\documentclass[prl,amsmath,amssymb,twocolumn,unsortedaddress]{revtex4-1}
\usepackage{graphicx}  % Include figure files
\usepackage{dcolumn}   % Align table columns on decimal point
\usepackage{bm}        % bold math 
\usepackage{float}
\usepackage{multirow}
\usepackage{amsmath}
\usepackage{amsmath,amssymb}
\usepackage{epsfig}
\usepackage{feynmp}
\usepackage{hyperref}
\usepackage{psfrag}
\usepackage[left=2.5cm,right=2.5cm,top=1.5cm,bottom=1.5cm,includefoot,includehead]{geometry}
\usepackage{color}

\newcommand{\lra}[1]{\langle #1 \rangle }
\newcommand{\bs}[1]{\boldsymbol{#1}}
\newcommand{\uu}[1]{{\bs{#1}}      }
\newcommand{\ds}[1]{\displaystyle{#1}}
\newcommand{\Om}{\Omega}
\newcommand{\R}{\mathbb{R}}
\newcommand{\OmB}{\Omega \setminus \bar{B}}
\newcommand{\D}{\mathbb{D}}
\newcommand{\vu}{\mathbf{u}}
\newcommand{\vf}{\mathbf{f}}
\newcommand{\vg}{\mathbf{g}}
\newcommand{\vF}{\mathbf{F}}
\newcommand{\vut}{\widetilde{\mathbf{u}}}
\newcommand{\vvit}{ \mathbf{v}}

\newcommand{\x}{\mathbf{x}}

\newcommand{\Frac}[2]  {\displaystyle {\frac{#1}{#2}}}
\newcommand{\dOm}{\partial \Om}
\newcommand{\dB}{\partial B}
\newcommand{\vtau}{\uu{\tau}}
\newcommand{\n}{\mathbf{n}}
\newcommand{\IntOm}[1]{\ds{\int_{\Om} #1}}
\newcommand{\IntBi}[1]{\ds{\int_{B_i} #1}}
\newcommand{\IntB}[1]{\ds{\int_{B} #1 }}
\newcommand{\IntPi}[1]{\ds{\int_{P_i} #1}}
\newcommand{\IntdBi}[1]{\ds{\int_{\partial B_i} #1 }}
\newcommand{\om}{\omega}

\newcommand{\Sum}[1]{\ds{\sum_{#1}}}
\newcommand{\Sumd}[2]{\ds{\sum_{#1}^{#2}}}
\newcommand{\stress}{\underline{\underline{\sigma}}}

\newcommand{\vue}{\mathbf{u}_{\epsilon}}

\newcommand{\eps}{ \varepsilon }

\newcommand{\Knab}{K_{\nabla}}
\newcommand{\HuzOm}{H^1_{0}(\Om)}
\newcommand{\KB}{K_{B}}
\newcommand{\LdOm}{L^2_0(\Om)}

\def\be{\begin{equation}}
\def\ee{\end{equation}}

\begin{document}
\setcounter{totalnumber}{10}

\title{Irreversibility and Chaos in  Active Particle Suspensions}

\author{Sergio~Chibbaro}
\affiliation{Sorbonne Université, Centre National de la Recherche Scientifique, UMR 7190, Institut Jean Le Rond d'Alembert,  F-75005 Paris, France}
\author{Astrid Decoene}
\affiliation{Universit\'e Paris Sud, Laboratoire de math\'ematiques d'Orsay (CNRS-UMR 8628), B\^atiment~425, 91405 Orsay cedex, France}
\email[]{astrid.decoene@math.orsay.fr}
\author{Sebastien Martin}
\affiliation{Universit\'e Paris Descartes, Laboratoire MAP5 (CNRS UMR 8145), 45 rue des Saints-P\`eres, 75270 Paris cedex 06, France}
\author{Fabien Vergnet}
\affiliation{Universit\'e Paris Sud, Laboratoire de math\'ematiques d'Orsay (CNRS-UMR 8628), B\^atiment~425, 91405 Orsay cedex, France}

\begin{abstract}
Active matter has been the object of huge amount of research in recent years for its important fundamental and applicative properties. In this paper we investigate active suspensions of micro-swimmers through direct numerical simulation, so that no approximation is made at the continuous level other than the numerical one. We consider both pusher and puller organisms, with a spherical or ellipsoidal shape.
We analyse the velocity and the characteristic scales for an homogeneous two-dimensional suspension and the effective viscosity under shear.
We bring evidences that the complex features displayed are related to a spontaneous breaking of the time-reversal symmetry. 
We show that chaos is not a key ingredient, whereas a large enough number of interacting particles and a non-spherical shape are needed to break the symmetry and are therefore at the basis of the phenomenology.
Our numerical study also shows that pullers display some collective motion, though with different characteristics from pushers.

\end{abstract}

\maketitle

\paragraph{Introduction}

Active matter is one of the major subjects of physical research because of its relevance in medicine, ecology and its possible applications~\cite{ramaswamy2010mechanics,bechinger2016active,koch2011collective,marchetti2013hydrodynamics,lauga2016bacterial,saintillan2018rheology}.
In nature, many living organisms swim through fluids at low Reynolds (Re) number~\cite{purcell1977life}.
 A particularly interesting instance of such low-Reynolds number world is given by a suspension of self-propelled particles, also called swimmers~\cite{pedley1992hydrodynamic,lauga2009hydrodynamics}, which are essential in life-cycle~\cite{munk1966abyssal} as well as in bio-engineering~\cite{dreyfus2005microscopic}. 
These small objects can exhibit complex dynamics as a result of the long-ranged hydrodynamic interactions which stem from the swimming activity, unfolding large-scale motion characterised by a various phenomenology of patterns, as highlighted in numerous experiments~~\cite{sokolov2009reduction,rafai2010effective,mino2011enhanced,cisneros2011dynamics,rusconi2014bacterial,petroff2015fast}. 
More specifically, bacteria may produce mesoscopic patterns of collective motion sometimes called “bio-turbulence”~\cite{dombrowski2004self,sokolov2007concentration,saintillan2012emergence,lopez2015turning}.
 In a flow, these bacteria may organize spatially and under shear they yield the possibility to increase or decrease the macroscopic viscosity to values above or below the suspending fluid viscosity, depending on their geometry and type of activity.
To understand the motion of such self-propelled particles, it has been analysed in numerical models, which are able to capture salient features of experiments, including hydrodynamic diffusion, large-scale collective motions and strong density fluctuations~\cite{hernandez2005transport,saintillan2008instabilities,drescher2011fluid,wensink2012meso,saintillan2018rheology,stenhammar2017role}.
Yet, different level of approximation either on the interactions or on the size of the objects are introduced to limit the computational cost, and the precise mechanisms underlying the phenomenology remain to be fully understood~\cite{sese2018velocity,fily2012athermal,caprini2019spontaneous}.
In this work, we develop a direct numerical approach that does not make use of any approximation other than considering a continuous medium,
and we focus on a striking feature of low-Re number flows, its reversibility in time~\cite{taylor1966low,purcell1977life}.
As emphasised in statistical mechanics since Boltzmann~\cite{cercignani1988boltzmann,boltzmann2012lectures,gallavotti2014nonequilibrium}, the relation between microscopic dynamics and macroscopic properties hides subtle issues, notably concerning the role of chaos and number of degrees of freedom~\cite{lebowitz1993boltzmann,chibbaro2014reductionism}.
It has been shown that non-Brownian passive suspensions break the time-reversal symmetry  at the level of single particles, even though they are governed by reversible creeping equations~\cite{pine2005chaos}. 
Identifying the mechanisms underlying  breaking of time-reversal symmetry in  biological swimmers and relate that to the macroscopic turbulent properties is the goal of this letter.
In particular, we bring evidence of a spontaneous breaking of the time-reversal symmetry, for which the key ingredients are: (i) a number of particles large enough to trigger important non-linear interactions among particles; (ii) an elongated shape of the particles.
It is important to underline here that if only the first condition is fulfilled, a complex ``bio-turbulent'' behaviour with large-scale collective motion is still encountered but the system remains reversible in time, showing no differences between spherical pushers and pullers and no particular rheological signature. 

\begin{figure}[h]
%\psfrag{FB}{$\mathbf F_{B_{i}}$}
%\psfrag{FP}{$\mathbf F_{P_{i}}$}
%\psfrag{FB}{$\mathbf F_{B_{i}}$}
%\psfrag{FP1}{$\mathbf F_{P_{i}}^{1}$}
%\psfrag{FP2}{$\mathbf F_{P_{i}}^{2}$}
\includegraphics[width=0.25\textwidth]{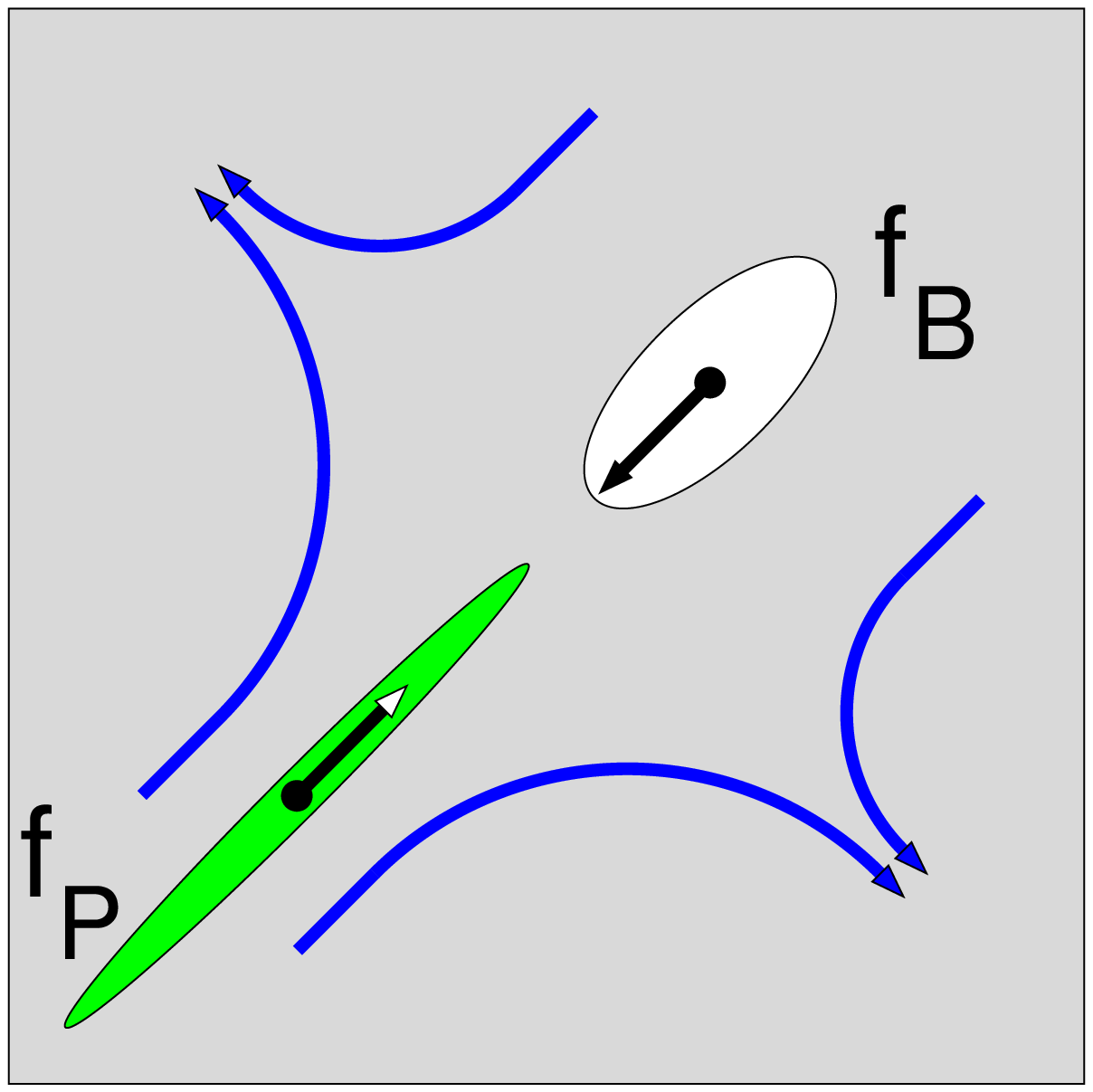}
\includegraphics[width=0.25\textwidth]{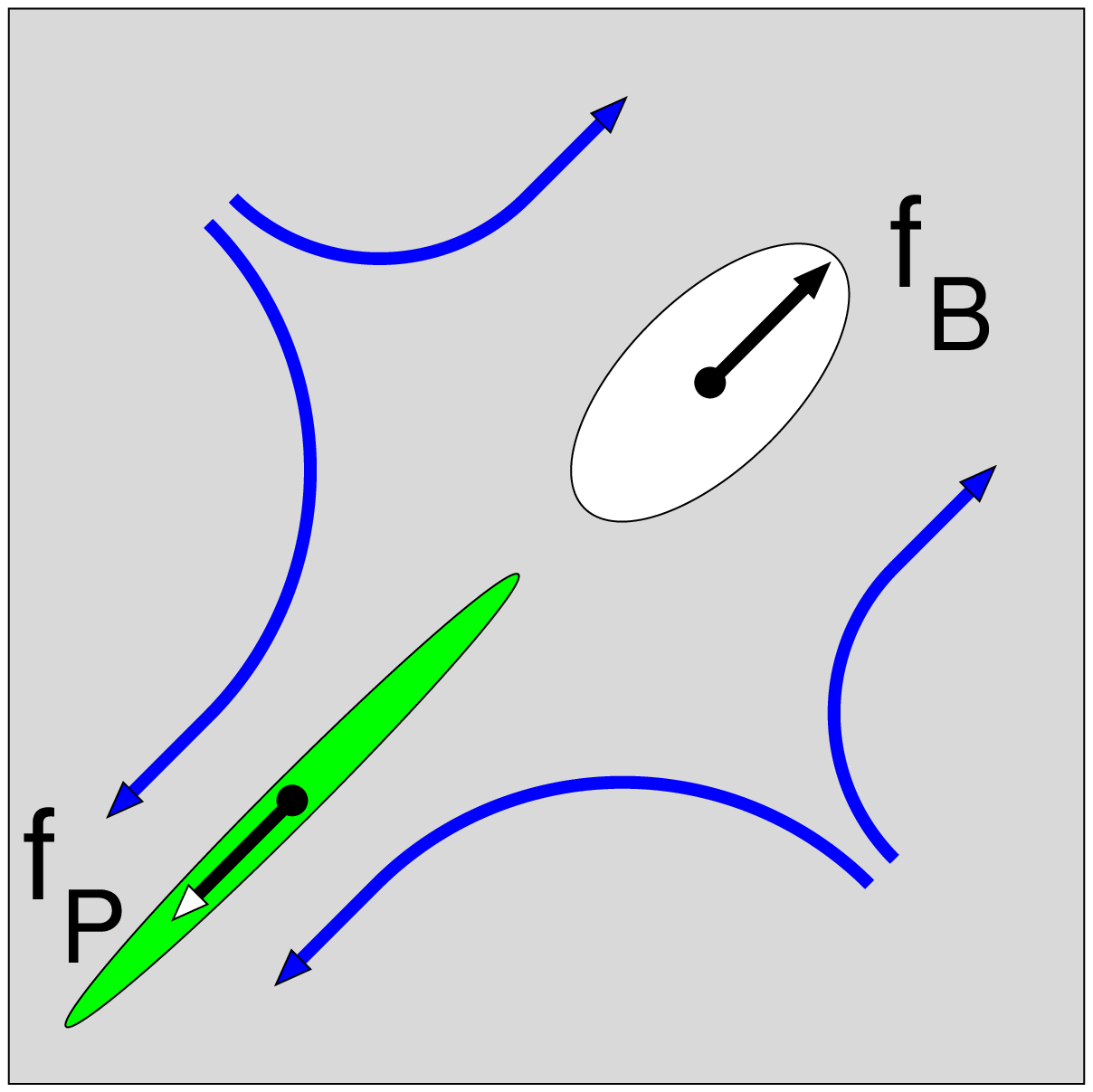}\\
\includegraphics[width=0.2\textwidth]{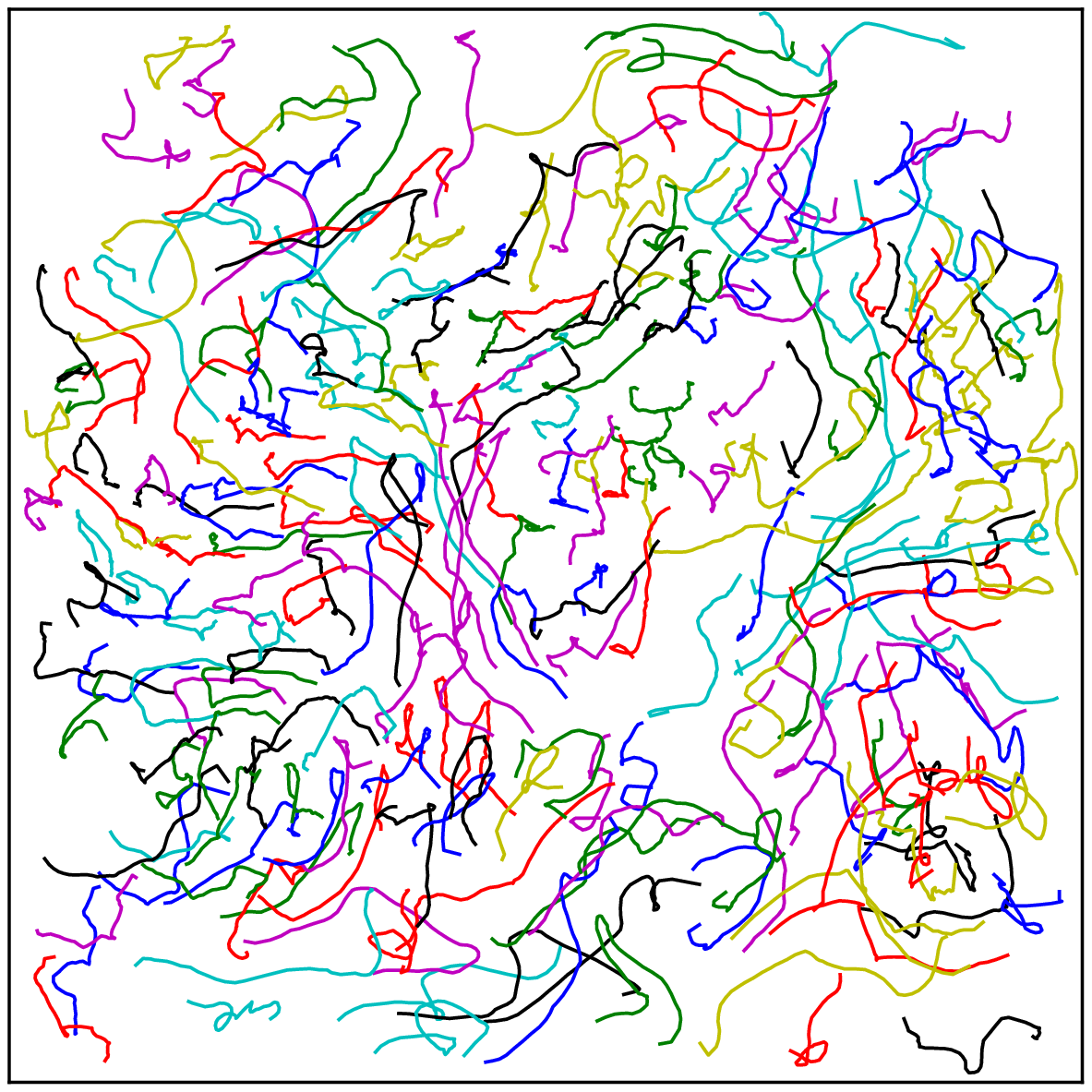}
\includegraphics[width=0.2\textwidth]{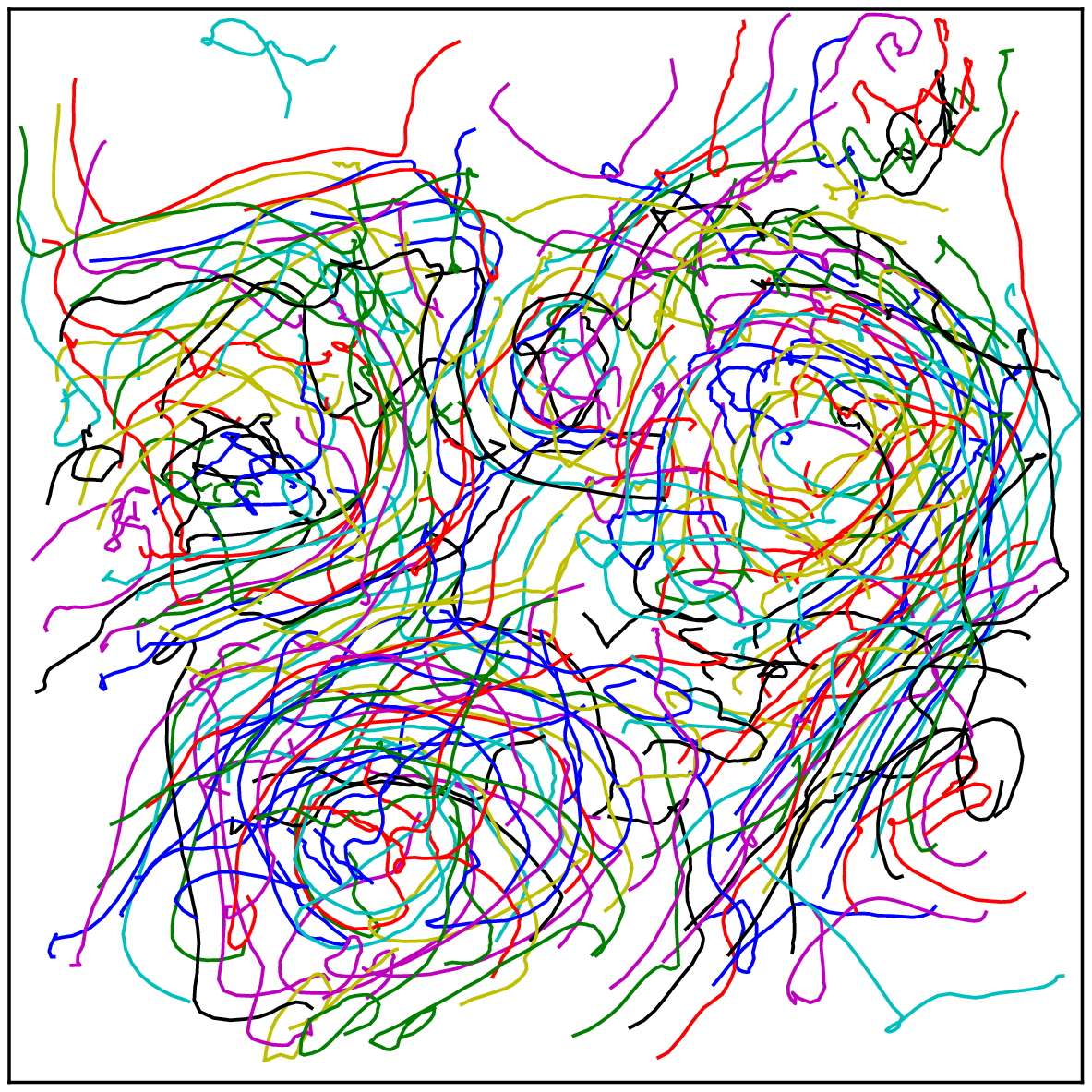}
\caption{
%Top: Domain $\Omega$ is made of a solid domain $B$ (collection of rigid bodies of the bacteria S) and a fluid (F) domain $\Omega \setminus B$.
Each bacterium is modeled by a rigid body $B$, with an associated propulsion force $\vf_B$, and the action of the flagellar bundle over the fluid is located inside a fluid region $P$ which can be deduced from the position of the rigid body.
The propulsion is such that the total force exerted by the swimmer on the fluid is equal to the total force of the fluid on the particle.
%The balance must be that the total force : 
%$\vF_P = - \vF_B.$
The propulsion force $\vf_B$ is directed outward from the center of $B$, parallel to the semi-major axes of $B$ and $P$, and has some orientation angle $\theta$.
The two types of swimmers are modeled in this way: {\em pushers} (right panels) push themselves forward by using flagella set back on the body;
%so that they generate a dipolar force field; 
{\em pullers} (left panels) pull themselves forward as they swim flagella first and generate the opposite flow field. More details on the model can be found in supplemental material~\cite{SUPP}.
Bottom: Swimmer displacements/trajectories in the homogeneous bi-periodic flow, after the transient period and with a volume fraction of $\phi=0.3$.}
\label{fig1}
\end{figure}

\paragraph{Theoretical Model}
 We model each micro-organism as a rigid ellipsoidal or spherical particle moving in the fluid. The flagella or cilia are not materialized; we only take into account the resultant force they exert on the particles and on the fluid. 
We deal with two kinds of flagellated swimmers, ``Pushers'', such as {\em Bacillus subtilis} or {\em Escherichia coli}, whose flagellar apparatus is localized at the back of the cell body, and ``pullers'' such as \emph{Chlamydomonas reinhardtii}, which typically propel themselves through flagella attached at the front of their cell body that execute a breaststroke-like motion~\cite{saintillan2018rheology}.
In fact, the complex movement of the locomotion structures (flagellar bundle or cilia) results in an  effective force acting on the fluid in a zone with non-zero volume, downstream (in the case of pushers), or upstream (in the case of pullers) of the organism. 
%Therefore we associate to each swimmer a dipole of forces, homogeneously distributed on the rigid body and on an elongated ellipsoidal region in the fluid representing the location of the flagella or cilia appendage. 
The modelled organisms are sketched in Fig. \ref{fig1}, where the model is presented.
%the swimmer's body is represented by an ellipsoid $B$ of constant semiaxis $R_B$ and $E_B$ (with $R_B \ge E_{B}$), and the force exerted by the flagellar bundle on the fluid, denoted by $\vF_P$, is supported in a very elongated ellipse $P$ of semiaxis $R_{P}$ and $E_P$ 
%(such that $R_P>R_B>E_{B}>E_{P}$),
%placed at a constant distance from the bacterial body. The propulsion force $\vF_B$ is directed outward from the center of $B$, parallel to the majorsemiaxes of $B$ and $P$, and has some orientation angle $\theta$.

A suspension of concentration $\phi$ is modeled by $N$ swimmers for the sake of simplicity in a two-dimensional domain $\Om =L^2$.
% filled with a Newtonian fluid governed by the Stokes equations. 
%The suspension has a volume concentration of $\phi$
%$= {|B|}/{|\Omega|}$, where $|\Omega|$ denotes the measure of a domain $\Omega$ . $
At the initial time the particles are distributed randomly over the fluid (without overlapping). The position of the center of the $i$th particle is denoted by $\x_{i}$, and by  $\vvit_{i}$ and $\om_{i}$ its translational and angular velocities.
We describe the fluid flow %in $\OmB$ 
by the Stokes equations :
%since inertial terms can be always neglected in the motion of microswimmers~\cite{Berg83,SKG07}:
\begin{equation} \label{prbm:ini:fluid}
\left\{\begin{array}{rcll}
-\mu \Delta \vu + \nabla p & = & \vf_{f} & 
%\mbox{in} \quad \OmB, 
\\
\nabla \cdot \vu & = & 0 & 
%\mbox{in} \quad \OmB,
\end{array}
\right.
\end{equation}
where $\vu=(u_{1},u_{2})$ and $p$ are the velocity and pressure field in the fluid.
This system is completed by specific boundary conditions: no-slip boundary conditions on the boundary $\dB_i$ of each rigid particle,
%\begin{equation} \label{cl:no-slip}
%\vu = \vvit_{i}+\om_{i} \times (\x-\x_{i})\  \textnormal{on $\dB_{i}$, $\forall i \in \{1,...,N\}$},
%\end{equation}
and 
%some Dirichlet or 
periodic conditions on $\dOm$, if not differently specified.
The forces considered on the fluid are the forces exerted by the locomotion structure.
%: $\vf_{f} = \Sum{i=1,...,N} \vf_{P_{i}} \chi_{P_{i}},$
%where $\chi_{P_{i}}$ is the characteristic function associated to the region  $P_{i}$ (and  $\chi_{B_{i}}$ wil be the characteristic function associated to the particle $B_{i}$). 
Newton's second law of motion governs the dynamics of the rigid bodies: force and torque balances are applied on each particle $B_i$ (see supplemental material for details).
%\begin{equation} \label{prbm:ini:part}
%%\left\{
%\begin{array}{rcl}
%\IntBi{\vf_{B_{i}}  } - \IntdBi{\stress \cdot \n } & = & 0  \quad \forall i \in \{1,...,N\},\\
%\IntdBi{  (\x-\x_{i}) \times  \stress \cdot \n } & = & 0  \quad \forall i \in \{1,...,N\},
%\end{array}
%%\right. 
%\end{equation}
%The fluid equations (\ref{prbm:ini:fluid}) and the particle equations (\ref{prbm:ini:part}) are coupled throughout the boundary  condition (\ref{cl:no-slip}).
The motion of each bacterium $B_{i}$ is then set by its instantaneous velocity,
$\vu(\x,t) = \vvit_{i}(t)+\om_{i}(t) \times (\x-\x_{i}(t))$
defined in $B_{i}$, and the dynamics by the differential equations:
$\dot{\x_{i}}(t) = \vvit_{i}(t), \quad \dot{\theta_{i}}(t) = \om_{i}(t)$.
We use random values for the initial data ${\x_{i}}(t=0)$ and ${\theta_{i}}(t=0)$.
The coupled fluid-particle problem is solved using the finite element method applied to solve the Stokes problem in the whole domain $\Omega$, and a penalty method to enforce the rigid motion constraint inside the rigid domain $B$. 
Possible numerical overlap between rigid particles due to the numerical errors must be prevented in order to guarantee robustness of the simulations. 
We have extended the numerical method previously proposed for granular gases~\cite{Mau06}, where inelastic collisions between rigid particles are computed.
%The equations are solved in non-dimensional units, so that results are given without dimensions. 
More details on the numerical method are given in the supplemental material~\cite{SUPP}.

%%%%%%%%%%%%%%
\paragraph{Numerical results}
In Fig. \ref{fig1}, we show the trajectories obtained by the simulation of  pushers and pullers with elongated ellipsoidal shape.
It is apparent that the dynamics is highly nontrivial in both cases, being characterised  by very irregular displacements which highlight the nonlinear interaction among swimmers which produce a chaotic motion. 
Yet, the two configurations show differences in the collective motion: pushers are able to build large rolls, as found also experimentally~\cite{cisneros2011dynamics}, while pullers appear to be coherent on smaller times, but are nonetheless found to be able to form collective rolls, though weaker than pushers, see~\cite{SUPP} for a movie. 
Swimmers tend therefore to form local inhomogeneities and to some extent synchronize their time trajectories, as found also in stochastic models~\cite{ten2011brownian,romanczuk2012active}.

%The particles in rolls are 
%previous computations based on a kinetic approach~\cite{saintillan2008instabilities}.
%These coherent vortical structures remind turbulent motion.
%{Similar features were found experimentally with other swimmers} like \emph{bacteria subtilis}~\cite{cisneros2011dynamics}.
%We investigate in the following the nature of this symmetry breaking.

\begin{figure}[h]
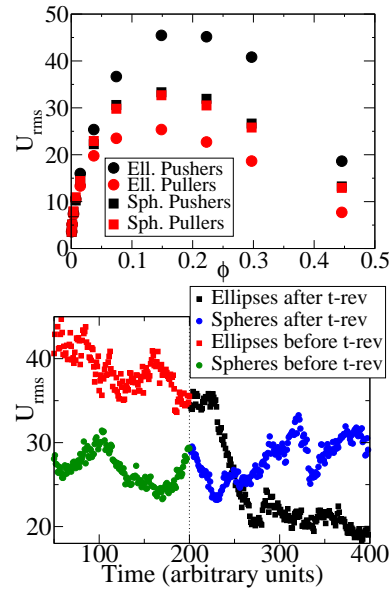

\includegraphics[width=0.3\textwidth]{fig2a.eps}
\includegraphics[width=0.3\textwidth]{fig2b.eps}
\caption{Top: Rms velocity defined as $U_{rms}\equiv \sqrt{\lra{U^2}}$  of the suspension, for pushers and pullers at different concentration, from dilute to dense suspensions. Both populations of swimmers have been studied with ellipsoidal  and spherical shape.
The velocity of the single particle is such that $U_{rms}\approx1$.
Bottom: We plot the root-mean-square velocity computed over all particles at each time-step for a suspension of pushers, both for ellipsoidal and spherical particles, for a concentration of $\phi=0.3$.
At the time $t_0=200$ the dynamics is inverted, which means $t\rightarrow -t$ and ${\bf V}\rightarrow -{\bf V}$.}
\label{fig2}
\end{figure}
To analyse quantitatively this issue, we have investigated the root mean square velocity of the organisms as displayed in Fig. \ref{fig2}.
It shows that the collective flow speed can be highly increased by hydrodynamic interactions in moderate to dense suspensions for both swimmers, and up to one order of magnitude for pushers consistent with experimental observations~\cite{SAKG07}.
Pushers are found to create stronger collective velocity than pullers, when the solid fraction is higher than few percents and elongated shape is taken into account. In dilute suspensions, swimmers are basically isolated and the dynamics is the same for all species.
The difference reaches a maximum at a concentration of around $20\%$, and then decreases at higher concentration due to congestion in dense suspensions, resulting in the close-pack structural configuration of the rigid micro-swimmers~\cite{cisneros2011dynamics}.
Remarkably, when the numerical experiment is performed with spheres the difference in the mean velocity is within the numerical errors.
Moreover, the mean velocity of the spheres is situated  between ellipsoidal pushers and pullers.
%\textcolor{red}{As shown in the supplemental material}, 
%The difference is traced back to the alignement of particles arising naturally from collective interactions, in accordance with experiments~\cite{cisneros2011dynamics}.
Looking at the velocity correlation, we have found that ellipsoidal pushers exhibit the most important local alignment, which explains the increase in velocity and larger coherent structures, whose typical length turns out to be about $0.3 L$.
Spherical pullers and pushers show the same properties, being reversible in time. In all cases some alignement is found, showing that pullers are also responsible for some structures even though of less importance \cite{SUPP}. This  differentiation suggests a time-symmetry breaking due to collective behaviour since applying the time-reversal operator to pullers one obtains formally the pusher dynamics and vice-versa. 
To highlight the issue, we have made the following experiment: we start a simulation with a pusher suspension, and at time $t_0$ we reverse the dynamics with a round-off error of $10^{-16}$. In principle, given that the system is reversible, it should retrace its steps.
It turns out that for ellipsoidal shape after a small amount of time in which the dynamics is reversed within numerical errors, the system breaks the time-symmetry, as shown in the Fig.\ref{fig2}b. Instead, for spherical active particles, the forward and backward time-trajectory is practically indistinguishable, at least from a statistical point of view.  

The irregular motion displayed in Fig. \ref{fig1} suggests also that swimmers are sensitive to small changes in the initial conditions,
that is they are chaotic in the sense of a positive Maximum Lyapunov exponent (LE)~\cite{ott2002chaos,vulpiani2010chaos}.
%, with a slight change in the relative positions of the particles producing a strong rearrangement. 
%\begin{table}[h]
%\begin{center}
%\begin{tabular}{c|rrr}
%$N$& $3$ & $15$ & $200$ \\
%\hline
%{pushers (ellips.)} &$-5.4$&$+26.2$&$-15.5$ \\
%{pushers (spheres)} &$-5.4$&$+26.2$&$-15.5$ \\
%{pullers (ellips)} &$+9.8$&$-32.9$&$+170.3$ \\
%{pullers (spheres)} &$+9.8$&$-32.9$&$+170.3$ \\
%\end{tabular}
%\end{center}
%\caption{The LE computed from the Euclidean distance between two simulations ${\Delta }(t)=\frac{1}{N} \sqrt{\sum_{i=1}^{N} (x_{i}^1-x_{i}^0)^2+(y_i^1-y_i^0)^2}$, where the two simulations are labelled with $1,0$, referring respectively to the perturbed and unperturbed case.
%We impose a perturbation $\delta {\bf x} (t=0)=10^{-10}$ on only one particle, because this should change all trajectories due to the hydrodynamical interactions.
%For chaotic systems, ${\Delta }$ grows exponentially as $\vert\delta {\bf x}\vert \exp(\lambda t)$, where $\lambda$ is the first Lyapunov exponent.
%To obtain smoother curves we average $\Delta$ over 50 different initial perturbations. The case with $N=3$ is studied when the particles are put initially aside as in~\cite{karolyi2000chaotic}.
%%If the particles are far initially, the system is found 
%}
%\label{tab1}
%\end{table}
Since in some cases the macroscopic irreversibility has been related to chaos~\cite{pine2005chaos,metzger2010irreversibility,metzger2013irreversibility}, 
we quantify this sensitivity computing the LE $\lambda$ for the two species of bacteria for different concentrations, both for ellipsoidal and spherical shape.
The LE is computed from the Euclidean distance between two simulations labelled $1,0$ as ${\Delta }(t)=\frac{1}{N} \sqrt{\sum_{i=1}^{N} (x_{i}^1-x_{i}^0)^2+(y_i^1-y_i^0)^2}$. 
%where the two simulations are labelled with $1,0$, referring respectively to the perturbed and unperturbed case.
%, as displayed in table \ref{tab1}.
The initial difference between the two simulations is set $\delta {\bf x} (t=0)=10^{-8}$ on only one particle,
%, because this should change all trajectories due to the hydrodynamical interactions.
and for chaotic systems, ${\Delta }$ grows exponentially as $\vert\delta {\bf x}\vert \exp(\lambda t)$.
%, where $\lambda$ is the first Lyapunov exponent.
%To obtain smoother curves we average $\Delta$ over 50 different initial perturbations. 
%The case with $N=3$ is studied when the particles are put initially aside as in~\cite{janosi1997chaotic,karolyi2000chaotic}.
%For sufficiently high volume fractions ($\phi>3\%$) both species are chaotic, whatever the shape, since the Lyapunov exponent is positive. 
All the cases are chaotic, except in the dilute regime where particles do not interact, and it is clear that shape does not play any role in chaos, as found also for passive particles~\cite{janosi1997chaotic,karolyi2000chaotic,pine2005chaos,metzger2010irreversibility}. Indeed, for a given concentration we have not measured any difference neither changing shape neither between puller and pusher. 
A  dependence on $\phi$ is yet found: $\lambda \approx 0.1\pm 0.01$ for $\phi \lesssim 0.2$ and $\lambda\approx 0.3\pm 0.02$ for $\phi\gtrsim 0.2$.
%\emph{Simulations reveal also a rapid increase in $\lambda$ with the volume fraction, which saturates above $\phi=?$}
Two mechanisms in principle can lead to such a chaotic behaviour, the N-body hydrodynamical interactions and the pair-contact ones. 
We have nevertheless verified that our results are insensitive to contacts, 
as already shown by the correct reversibility of spherical objects.
We also assure a such small time-step that particles do not reach each other except in very rare cases. {Furthermore, the amplification is almost the same even without taking into account the contacts, as already shown in sheared suspensions}~\cite{metzger2013irreversibility}. 
Hence, long-range hydrodynamic interactions lead to a chaotic regime, but chaos does not play any role in the macroscopic irreversibility.

\begin{figure}[h]
\includegraphics[width=0.3\textwidth]{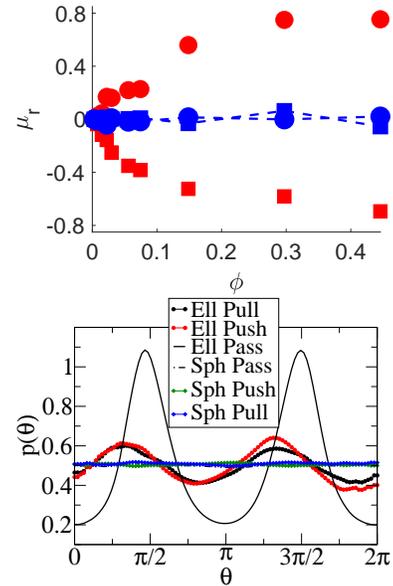}
\includegraphics[width=0.3\textwidth]{fig3b.eps}
\caption{(a) Evolution with the concentration in a suspension of ellipsoidal pushers and pullers of the relative effective viscosity $\mu_r=(\mu-\mu_{\text{passive}})/\mu_{\text{passive}}$, where $\mu_{\text{passive}}$ is the effective viscosity of the suspension for non-active particles.
Pushers are indicated by the symbol $\square$, and Pullers by $\bullet$. Red symbols indicate elongated particles, while blue-dashed ones are for spherical particles.
(b) Probability density function of orientations in the simulations of pusher, puller and passive suspensions. The concentration is $\phi=0.15$ and the ellipsoidal particles are of elongation ratio $2$.
}
\label{fig3}
\end{figure}
We analyse now some of the rheology characteristics of swimmer suspensions.
We focus on the study of the shear-viscosity in a fluid confined between two parallel rigid plates with a steady relative shear motion. In this case one can measure the shear stress, which is defined as the average force applied by the fluid on the plates per surface unit in response to this shear
$ F = {\int_\Gamma (\bf{\sigma} \cdot \n) \cdot \uu{\tau}}/{2L}$n
where $\Gamma$ denotes the surface of the two plates, $L$ denotes the length of each plate, $ \n$ is the normal vector $\Gamma$ pointing outward and $\uu{\tau}$ is the tangential vector on $\Gamma$ opposed to the shear flow.
Consequently, for a suspension, an apparent viscosity is defined as 
$\mu_{\mathrm{app}}(t)={F}/{\dot{\gamma}}$, where $\dot{\gamma}$ denotes the shear rate. Since this value changes in time due to the evolution of the particles configuration, one usually considers the effective viscosity:
$
\mu_{\mathrm{eff}}= \lim_{T \mapsto + \infty} \frac{1}{T} \int_0^T  \mu_{\mathrm{app}}(t)\, \mathrm{d}t
$.
We have computed the effective viscosity in active suspensions of pusher- and puller-like swimmers using 2D simulations in which shear is imposed through non-homogenous Dirichlet conditions at walls and periodic boundary conditions are imposed on the left and right boundaries.
It is well known that the viscosity of a passive suspension of particles is different from that of the solvent \cite{einstein1906einenenu,einstein1911berichtigung,batchelor1977effect}, and that shape of the particles has an effect \cite{brenner1974rheology}.
We analyse here the impact of the active (swimming) motion looking at the effective relative viscosity, that is eliminating the rheology signature related to the passive suspension, see Fig.~\ref{fig3}.
Beyond the dilute regime, elongated pusher-like swimmers tend to decrease the effective viscosity, while pullers increase it~\cite{GLAB11}. Yet spherical ones have no relative rheological signature, spherical active particles behaving like the passive ones~\cite{HABK08,IP07a,saintillan2018rheology}.
Furthermore, puller and pushers act in a symmetrical way with respect to the corresponding {spherical} suspensions, as highlighted by Fig. \ref{fig3}.
As shown in Fig. \ref{fig3}b, physically, the broken symmetry by elongation translates into 
a preferential alignment in the flow at positions which maximise (minimise) the effective viscosity for  Pullers (Pushers).
Passive ellipses have preferential alignment but symmetrical with respect to the neutral position. Spheres are not able to break the symmetry and do not show any preferential alignment.
See the supplemental material for more details.
Our study shows again that, if no other mechanism is added, elongation is needed to produce an additional rheological signature,
which points out again the link to  
spontaneous symmetry breaking by shape.
%As displayed in Fig. \ref{fig3}b,  the symmetry with respect to the orientation is broken in the case of elongated particles. 
%In fact, the latter tend to align in the flow at orientation $\pi/4$ (or equivalently $5\pi/4$), due to interactions with the other particles. Since active particles align in an orientation that maximizes apparent viscosity for puller-like swimmers, and minimizes it for pusher-like swimmers, effective viscosity is enhanced in the first case, and diminished in the second case.
%Spheres are not able to break the symmetry, and while they also align preferentially at two positions $\theta=\pi/2 , 3 \pi /2$,  
%they stay the same amount of time at these two symmetrical positions, and rheological contribution hence averages to zero. 
%Furthermore, our results show that steric interactions are not responsible for the collective properties of micro-swimmers.
%We have then investigated the diffusivity properties of the system looking at the mean square displacement $\lra{\vert\Delta {\bf x}\vert ^2}$ averaged over all the particles and many realizations. This turns out to increase linearly with time, after a short transient in which it is ballistic. We use these data to compute the effective diffusivity $D=\lra{\vert\Delta {\bf x}\vert ^2}/2t $, see figure \ref{fig2}b. \textbf{(Risultati strani da rivedere mancano troppo densita')}
\paragraph{Conclusions}
We have reported on numerical experiments of swimmer dynamics obtained without other approximations than the hydrodynamics, the fluid-structure interaction being explicitly accounted for. 
%Swimmers of pusher and pullers kind are studied changing the force generated by the flagella.
The simulations are able to reproduce all the relevant dynamical features. We have shown unambiguously that all the features characteristic of the collective bio-turbulent regime may be related to the following ingredients: the number of active particles and the elongation.
The former is necessary to trigger interactions between velocities which in turn provoke a chaotic motion. In particular, alignement of particles are obtained, even without explicit interaction between directions. Elongation allows to spontaneously break the symmetry under time-reversal which explains the observed impact both on motion and rheology. Interestingly, Pullers are found to produce "bio-turbulence", even though of less entity than pushers.
Simplified models and artificial devices could be based upon these conclusions.

\bibliographystyle{abbrv}
\bibliography{references}

%%%%%

\clearpage

\onecolumngrid

\renewcommand{\thesection}{S\arabic{section}}
\renewcommand{\thefigure}{S\arabic{figure}}
\renewcommand{\theequation}{S\arabic{equation}}
\renewcommand{\thetable}{S\arabic{table}}

\section*{SUPPLEMENTAL MATERIAL}

\section {Model}

%``Pushers'' are swimmers whose flagellar apparatus is located at the back of the cell body. The flagella generate a thrust force at the back which is cancelled by the drag force on the cell body in the front. In contrast, ``pullers'' typically propel themselves through flagella attached at the front of their cell body that execute a breaststroke-like motion. We model each micro-organism as a rigid ellipsoidal particle moving in the fluid. The flagella are not materialized; we only take into account the resultant force they exert on the particles and on the fluid. In fact, the complex movement of the locomotion structures results in an  effective force acting on the fluid in a zone with non-zero volume, downstream (in the case of pushers), or upstream (in the case of pullers) of the organism. . %Expliquer mieux: il s'agit d'une moyenne sur un certain temps (au moins dans le cas des cils).

\begin{figure}[h]
\psfrag{S}{Solid domain}
\psfrag{F}{Fluid domain}
\psfrag{X1}{$x_{i}$}
\psfrag{X2}{$y_{i}$}
\psfrag{B}{$B_{i}$}
\psfrag{P}{$P_{i}$}
\psfrag{T}{$\theta_{i}$}
\centerline{A)\includegraphics[scale=0.45]{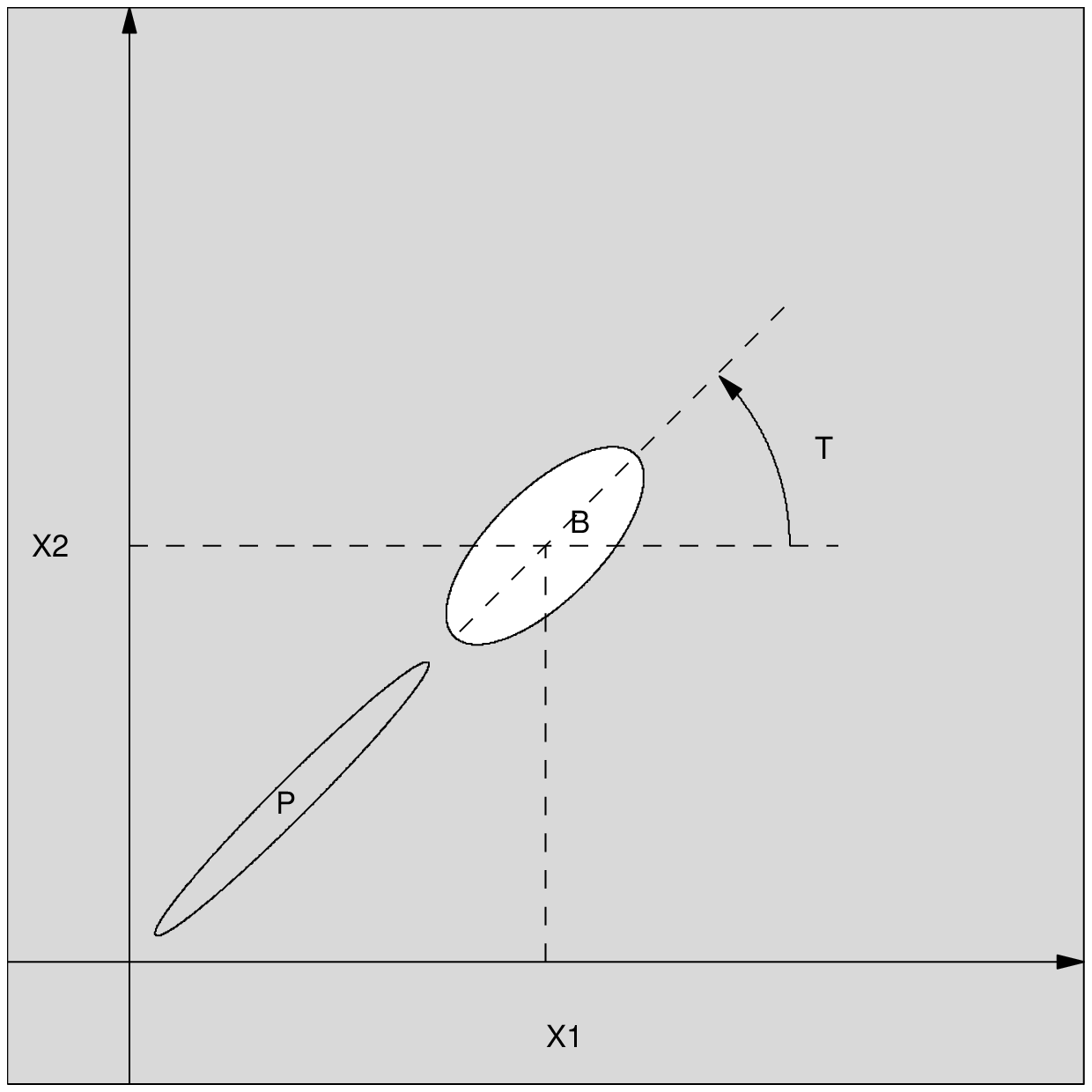}
B)\includegraphics[scale=0.45]{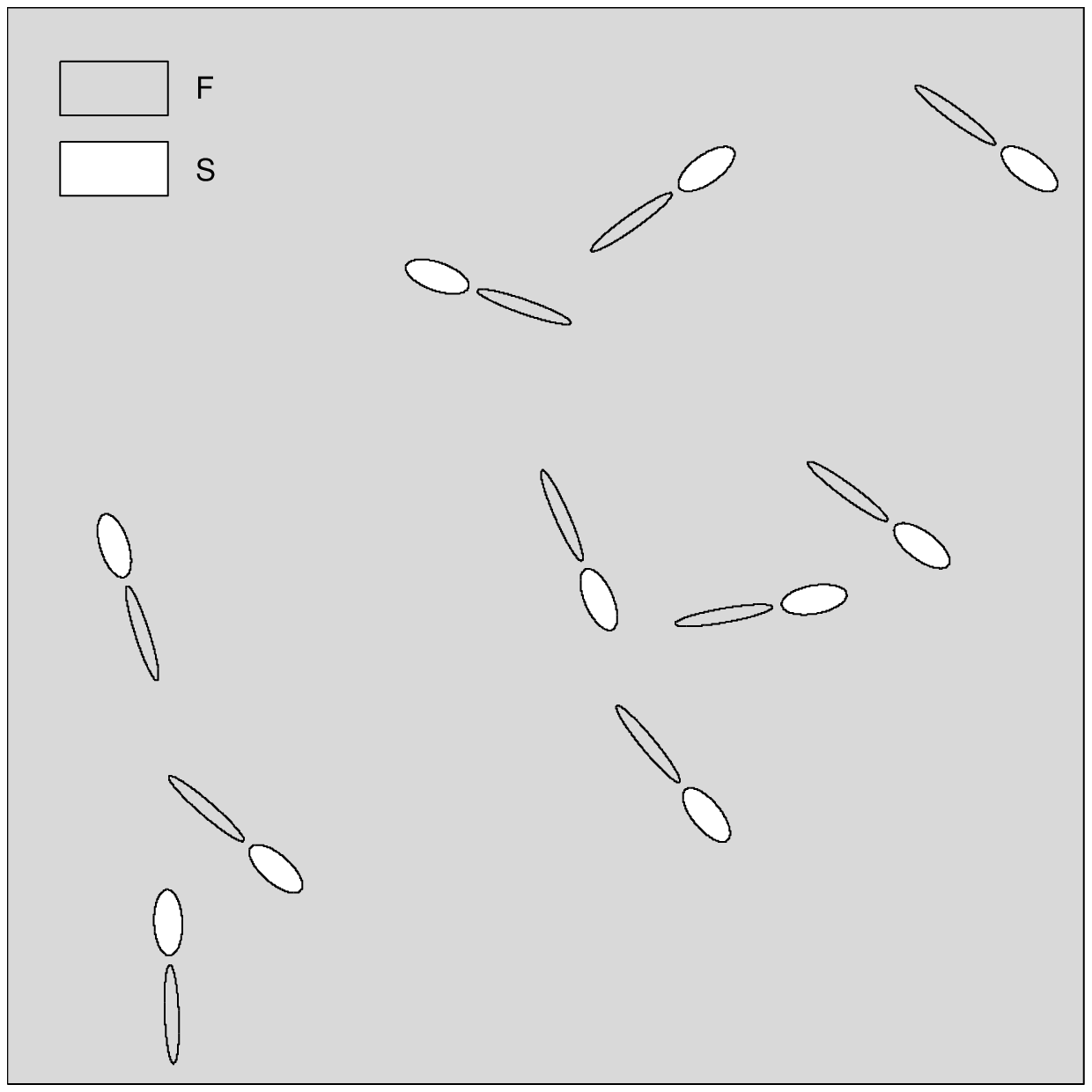}}
\caption{A) Domain $\Omega$ is made of a solid domain $B$ (collection of rigid bodies of the bacteria) and a fluid domain $\Omega \setminus B$. 
Denoting by $(B_{i})_{i=1,...,N}$ the rigid particles representing the body of the swimmers, strongly included in $\Om$, and by $B=\cup_i B_{i}$ the whole rigid domain, the suspension has a volume concentration of $\phi= {|B|}/{|\Omega|}$, where $|\omega|$ denotes the measure of a domain $\omega$ . 
B) Each bacterium is modeled by a rigid body $B_{i}$. Its position is characterized by the coordinates $\mathbf x_{i}=(x_{i},y_{i})$ of its center of mass and its orientation $\theta_{i}$. The action of the flagellar bundle over the fluid is located at a fluid region $P_{i}$ which can be deduced from the position of the rigid body.}\label{fig:domain}
\end{figure}

In this section, we briefly describe the continuous model used. 
Each swimmer's body is represented by a rigid ellipsoid $B$, to which we associate a dipole of forces, homogeneously distributed inside the rigid body and inside an elongated ellipsoidal region $P$ in the fluid, placed at a constant distance from the bacterial body and representing the location of the flagella appendage, see Fig.~\ref{fig:domain}.
%A suspension is modeled by $N$ swimmers in a two-dimensional domain $\Om \subset \R^2$ filled with a Newtonian fluid governed by the Stokes equations. 
%At the initial time the particles are distributed randomly over the fluid (without overlapping). 
%The position of the center of the $i$th particle is denoted by $\x_{i}$, and by  $\vvit_{i}$ and $\om_{i}$ its translational and angular velocities.
The propulsion force exerted on the body of each swimmer $i$ is denoted by $\vF_{B_i}$, has constant magnitude $f_P$ and is directed outward from the center of $B_i$, parallel to the majorsemiaxes of $B_i$ and $P_i$. It has some orientation angle $\theta_i$, as shown in Fig.~\ref{fig:domain}(A), and satisfies :
$$
\vF_{B_{i}} = f_{P}  \vtau_{i} = \IntBi{ \vf_{B_{i}} \, \mathrm{d}\x }, \quad \mbox{with}  \quad \vf_{B_{i}} = \frac{f_{P}}{|B_{i}|} \vtau_{i},
$$
where $\vtau_{i}$ is given by the orientation angle $\theta_{i}$. 
Since each swimmer is force and torque-free, the total force exerted by the swimmer on the fluid is equal to $-\vF_{B_i}$, and we can write:
$$
\vF_{P_{i}}  = - f_{P} \vtau_{i} = \IntPi{ \vf_{P_{i}} \, \mathrm{d}\x }, \quad \mbox{with}  \quad \vf_{P_{i}} = -  \frac{f_{P}}{|P_{i}|} \vtau_{i}.
$$
%the unit vector orientated along the direction  $\theta_{i}$. 

%
%\begin{figure}
%\psfrag{FB}{$\mathbf F_{B_{i}}$}
%\psfrag{FP}{$\mathbf F_{P_{i}}$}
%\psfrag{FB}{$\mathbf F_{B_{i}}$}
%\psfrag{FP1}{$\mathbf F_{P_{i}}^{1}$}
%\psfrag{FP2}{$\mathbf F_{P_{i}}^{2}$}
%\centerline{A)\includegraphics[scale=0.35]{NEW_FIGURES/fig_JFM_Pusher_dessin.eps}B)\includegraphics[scale=0.35]{NEW_FIGURES/fig_JFM_Puller_dessin.eps}}%\includegraphics[scale=0.35]{NEW_FIGURES/fig_JFM_Chlamy_dessin.eps}}
%\caption{Individual models of microswimmers. A) a pusher is represented by a rigid body and a flagellar area. B) A puller can be modeled by inverting the propulsion force.}%C) More realistic models can be designed, such as {\em Chlamydomonas reinhardtii}: the action of the cilia over the fluid is modeled by the total force $\mathbf F_{P_{i}}=\mathbf F_{P_{i}}^1+\mathbf F_{P_{i}}^2$.
%\label{fig:swimmer-model}
%\end{figure}

The Reynolds number is typically less than $10^{-2}$  \cite{Berg83,SKG07}, therefore inertial effects can be neglected and the fluid flow is described by the Stokes equations :
\begin{equation} \label{prbm:ini:fluid}
\left\{\begin{array}{rcll}
- \nabla \cdot \underline{\underline{\sigma}}(\vu,p) & = & \vf_{f} & \mbox{in} \quad \OmB, \\
\nabla \cdot \vu & = & 0 & \mbox{in} \quad \OmB,
\end{array}
\right.
\end{equation}
where $\vu$ and $p$ are the velocity and pressure field in the fluid, and where the stress tensor $\underline{\underline{\sigma}}$ writes 
$$\underline{\underline{\sigma}}  =  2 \mu \D(\vu) -  p\, \mathbb{I}, \quad \mbox{where} \quad \D(\vu) = \Frac{\nabla \vu + (\nabla \vu)^T}{2}$$
since we consider a Newtonian fluid. This system is completed by specific boundary conditions: no-slip boundary conditions on the boundary $\dB_i$ of each rigid particle
and some Dirichlet or periodic conditions on $\dOm$. The forces considered on the fluid are the forces exerted by the flagella:
$$\vf_{f} = \Sum{i=1,...,N} \vf_{P_{i}} \chi_{P_{i}},$$
where $\chi_{P_{i}}$ is the characteristic function associated to the region  $P_{i}$ and  $\chi_{B_{i}}$ is the characteristic function associated to the particle $B_{i}$. Finally, 
%Newton's second law of motion governs the dynamics of the rigid bodies:
force and torque balance for each body writes:
\begin{equation} \label{prbm:ini:part}
\left\{
\begin{array}{rcl}
 \IntBi{{\rho} \vg} + \IntBi{\vf_{B_{i}}  } - \IntdBi{\stress \cdot \n } & = & 0  \quad \forall i \in \{1,...,N\},\\
\IntdBi{  (\x-\x_{i}) \times  \stress \cdot \n } & = & 0  \quad \forall i \in \{1,...,N\},
\end{array}
\right. 
\end{equation}
where $\rho$ denotes  the buoyant density of swimmers, which is positive since these are slightly denser than the fluid, and $\vg=(0,-g)$ is gravity. 

The motion of each bacterium $B_{i}$ is then set by its instantaneous velocity:
$$\vu(\x,t) = \vvit_{i}(t)+\om_{i}(t) \times (\x-\x_{i}(t))$$ 
defined on $\dB_{i}$. More precisely, the individual translational velocity is equal to the average velocity:
\begin{equation} \label{translVel}
\vvit_{i}(t) = \frac{1}{| \dB_{i}(t) |} \IntdBi{ \vu(t,\x)\, \mathrm d\x} ,
\end{equation}
and the individual angular velocity is equal to
\begin{equation}\label{angVel}
\om_{i}(t) = \frac{1}{| \dB_{i}(t) | } \displaystyle \frac{\IntdBi{ \vu(t,\x) \times (\x-\x_i(t))\, \mathrm d\x}}{\IntdBi{ |\x-\x_i(t)|^2\, \mathrm d\x}}.
\end{equation}

The swimmers dynamics are then set by the differential equations:
\begin{equation}\label{dyn}
\dot{\x_{i}}(t) = \vvit_{i}(t) \, , \qquad \dot{\theta_{i}}(t) = \om_{i}(t).
\end{equation}
These introduce a dependency in time into the problem: Stokes equations are steady, but their solution depends on time because the configuration of the swimmers inside the fluid varies in time.\\

\section{Numerical scheme}

The coupled fluid-particle problem is solved using a new specific finite element method applied to solve the Stokes problem in the whole domain $\Omega$, and a penalty method to enforce the rigid motion constraint inside the rigid domain $B$. This method is rigorous in the sense that the numerical solution mathematically converges to the solution of the fluid-structure problem as the penalty parameter tends to $0$.
%Possible numerical overlap between rigid particles due to errors in the solution of the Stokes equations must be prevented in order to guarantee robustness of the simulations. We have extended the numerical method proposed in~\cite{Mau06}, where inelastic collisions between rigid particles are computed. Denoting by $\Delta t > 0$ the time step, the fluid-particle problem (\ref{prbm:ini:fluid})-(\ref{prbm:ini:part}) is solved at each time step $t=t^n=n\Delta t$, allowing to compute an approximation of the translational and angular velocities $\vvit_{i}^n$ and $\om_{i}^n$ for each swimmer. 
The time algorithm reads as follows: in a first step, the fluid-particle problem is solved without taking into account the possible overlapping of the particles (thus defining an {\it a priori} velocity of the swimmers), then the projection of this {\it a priori} velocity onto the set of admissible velocities is computed. Finally, the position and orientation of each swimmer is updated.
The position and orientation of each swimmer at time $t^{n+1}$ are updated by solving the ODEs (\ref{dyn}) using for instance a second-order Adam-Bashfort method.
A brief description of the schemes are given in the following.

\subsection*{A fictitious domain approach}

%Our purpose is to avoid mesh generation of the moving and complex fluid domain,
%: the complexity of the mesh due to the presence of inclusions would no allow to use standard fast solvers, and the computational cost of remeshing could become prohibitive. 
%therefore we choose a 
The fictitious domain approach we use allows to avoid remeshing. %One possibility is to use an iterative algorithm on an auxiliary field, composed by Lagrange multipliers, which warrants the rigid motion constraint of the particles (see for instance~\cite{GLO01,GLO03}). 
%An alternative to this are 
We use a penalty method: the rigid motion constraint is obtained by relaxing a term in the variational formulation, what amounts to replace rigid zones by highly viscous ones (see~\cite{Ja05,Le07,Calta04}). 
%In this work we choose a penalty method already used in \cite{Le07} for the simulation of passive particles in a fluid.\\
The mathematical problem is solved in the 
%Let us first introduce the 
following constrained functional spaces:
$$
\begin{array}{rcl}
\Knab \ = \ \left\{ \vu \in \HuzOm, \ \nabla \cdot \vu = 0 \right\}, \quad 
\KB \ = \ \left\{ \vu \in \HuzOm, \ \D(\vu) = 0  \ \textrm{a.e. in}\ B  \right\}.\\
\end{array}
$$
$\Knab$ is the space of divergence free functions defined on $\Om$ and $\KB$ is the space of functions which do not deform $B$. The solution to the initial problem, defined on $\OmB$, can be extended on the whole domain $\Om$ by a function in $\KB$: 
$ \vu(\x,\cdot)=\vvit_{i} + \om_{i} \times ( \x-\x_{i}) $ in $B_{i}$ for every $i$, and we still denote this extension by $\vu$. 
The problem in variational form can then be written as the minimization of the functional 
$$
J(\vu) \ = \  \mu \, \IntOm |  \D(\vu) |^2  \ - \  \IntOm \vf \cdot \vu
$$
on $\KB \cap \Knab$,
%} =  \left\{ \vu \in \HuzOm \ /  \ \grad \cdot \vu = 0, \ \red{\D(\vu) = 0}  \ p.p. \ dans \ B   \right\}$$
%avec
%$$
%\vf \ = \ \Sumd{i=1}{N} \left(  \vf_{b}^i  \ \chi_{B_{i}} \ + \ \vf_{p}^i  \ \chi_{P_{i}} \right) \ .   
%$$\\
where 
$$
\vf = \Sumd{i=1}{N} \left( \vf_{b}^i \, \chi_b^i +  \vf_{p}^i  \, \chi_p^i \right).$$
%and $\LdOmD$ stands for the set of $L^2$ functions over $\Omega$ with zero mean value and equal to $0$ inside $B$.
The rigid motion constraint is relaxed by introducing the following penalty term in the functional to minimize:
$$
\int_{B} \frac{1}{\eps}\ \D(\vu) \, : \, \D(\vu),
$$
so that $\D(\vu)$ goes to zero in $B$ when $\eps$ goes to zero and $\vu$ tends to a rigid motion in $B$.
The variational formulation obtained is: find $\vue \in \HuzOm$ and $p \in \LdOm$ such that
\begin{equation}\left\{
\begin{array}{rcll}\label{varform:penal}
%
%\vspace{3.mm}
2 \mu \, \IntOm  \D(\vue) \, : \,  \D(\vut) 
 + \frac{2}{\eps} \,  \IntB  \D(\vue) \, : \,  \D(\vut) 
-  \IntOm {p_{\epsilon}} \, \nabla \cdot \vut  & = & \IntOm \vf \cdot \vut, &  \forall \vut \in \HuzOm, \vspace{3mm}\\
%
%& & \\
% 
\IntOm q \, \nabla \cdot \vue & = & 0, & \forall q \in \LdOm,
\end{array}
\right.
\end{equation}
It has been proven in \cite{Ja05,maurySiam} that the penalty method converges linearly in $\eps$.
%: the solution $\vue$ of problem (\ref{varform:penal}) converges to the solution $\vu$ of the initial problem as $\eps$ vanishes, and the convergence is of order $1$ in $\eps$.
%We refer to~\cite{maurySiam} for a detailed analysis of a scalar version of this problem, which provides an error estimate for the space-discretized problem at the order $\eps + h^{1/2}$.
%$\mathcal{O} (\epsilon) +  \mathcal{O} (h^{1/2})$.
%However, this method is not applicable in $3$ dimensions, because the penalty term strongly degrades the condition number of the matrix that has to be inverted at each time step (and which comes from the penalized Stokes problem). An iterative method would therefore converge too slowly for solving the associated linear system, and since direct methods involve the storage of matrices their use is not reasonable in $3$ dimensions. Other fictitious domain methods must be used if one wishes to make direct simulations of active suspensions in $3$d, like for example the method proposed in \cite{BFa} for passive suspensions. Another possibility is to use a suitable preconditioner as in \cite{ChPo}.

\subsection*{Contact algorithm}

In the present hydrodynamic framework, it is known that contacts are not supposed to happen (see~\cite{GVH10,H07}).
Yet in actual simulations, collisions between particles are likely to  occur because of the numerical error. 
%From a numerical point of view, it means that the particles may overlap when their positions are updated after the velocity field computation. 
The treatment of possible overlaps is crucial in the case of dense suspensions.
% if we want to avoid small time steps. 
To address this issue, we have extended to wet ellipses 
the numerical method already proposed for dry spheres~\cite{Mau06}.
% proposed by Maury
%introduced in~\cite{Mau06} 
%in the context of dry granular flows, with inelastic collisions between grains. We propose to integrate this approach in the wet situation. The interest in the procedure relies on the possibility to use any suitable solver for the computation of the dynamics. Contacts are handled at a second stage, without any consideration of the proper dynamics. 
%where inelastic collisions between rigid particles are computed. 
The method consists of projecting the velocity field onto some convex admissible set depending on the current configuration, so that particles do not overlap.
%In this work, the contact algorithm has been extended to the case of ellipsoidal particles. 
Since no analytical expression of the minimal distance between two ellipses is available, an approximation of this distance must be computed at each time, by solving the problem of searching for the proximal points on a each pair of ellipsoidal particles. 

Let us detail the method in the case of spherical particles:
we denote by $\mathbf{X}^n:=(\mathbf x_{i}^n)_{i=1,...,N}$ the position of $N$ particles (more precisely, the position of their gravity centre) at time $t_{n}$, and by 
$\hat{\mathbf{V}}^{n}$ their \emph{a priori} velocity.
%Similarly, we denote by $\vtheta^n:=(\theta_{i}^n)_{i=1,...,N}$ the set of angular orientations of the bacteria (this allows us to locate the flagella) at time $t_{n}$. 
We define the set
\begin{equation}
\label{eq:defK}
K(\mathbf{X}^n) =\left\{ \mathbf{V} \in \mathbb{R}^{2N},\ D_{ij}(\mathbf{X}^n) + \Delta t\, \mathbf{G}_{ij}(\mathbf{X}^n)\cdot \mathbf{V} \geq 0,\, \forall i < j \right\},
\end{equation}
where $$D_{ij}(\mathbf{X}^n)=\left| \mathbf{x}_{i}^n-\mathbf{x}_{j}^n\right| -2 R_{b}$$ denotes the signed distance between two spheres $B_i$ and $B_{j}$ and $$\mathbf{G}_{ij}(\mathbf{X}^n)=\nabla D_{ij}(\mathbf{X}^n)=(...,0,-\mathbf{e}_{ij}^n,0,...,0,\mathbf{e}_{ij}^n,0,...),\qquad \mathbf{e}_{ij}^n=\displaystyle \frac{\mathbf{x}_j^n-\mathbf{x}_i^n}{\left| \mathbf{x}_j^n-\mathbf{x}_i^n\right|}$$ is the gradient of the distance. 
%
%Our purpose is to solve the dynamics problem by constraining the velocity field of the particles to belong to the set of admissible velocity fields, at each time step, in order to prevent the particles from overlapping.
%\begin{rmrk}
%Denote $E(\mathbf{X}^n) =\left\{ \mathbf{V} \in \mathbb{R}^{2N},\ D_{ij}(\mathbf{X}^n+ \Delta t\, \mathbf{V}) \geq 0,\ \forall i < j \right\}$ the set of velocity fields $\mathbf{V}$ such that particles, at position $\mathbf{X}^n$ at time $t_n$ and with velocity $\mathbf{V}$, do not overlap at the next time step. The constraint $D_{ij}(\mathbf{X}^n) + \Delta t\, \mathbf{G}_{ij}(\mathbf{X}^n)\cdot \mathbf{V} \geq 0$ is the linearized form of the constraint $D_{ij}(\mathbf{X}^n+ \Delta t\, \mathbf{V}) \geq 0$ and, furthermore, it can be shown that $K(\mathbf{X}^n) \subset E(\mathbf{X}^n)$. It means in particular that particles with admissible velocities at time~$t_n$ do not overlap at time $t_{n+1}$.
%\end{rmrk}
%
In order to avoid overlapping, the following splitting procedure is proposed: in a first step, we solve the variational problem without taking into account the possible overlapping of the particles (thus defining an {\it a priori} velocity of the spheres), then compute the projection of this {\it a priori} velocity onto the set of admissible velocities defined by~(\ref{eq:defK}). The constrained problem is formulated as a saddle-point problem, by using the introduction of Lagrange multipliers:
$$\left\{\begin{array}{l}
\textnormal{Find $(\mathbf{V}^{n},\mathbf{\Lambda}^{n})\in \mathbb{R}^{2N} \times \mathbb{R}_{+}^{N(N-1)/2}$ such that}\\
\mathcal{J}(\mathbf{V}^{n},\mathbf{\lambda}) \leq \mathcal{J}(\mathbf{V}^{n},\mathbf{\Lambda}^{n}) \leq \mathcal{J}(\mathbf{V},\mathbf{\Lambda}^{n}),\quad \forall (\mathbf{V}, \mathbf{\lambda}) \in \mathbb{R}^{2N} \times \mathbb{R}_{+}^{N(N-1)/2},			
\end{array}\right.$$
with the following functional:
$$
\mathcal{J}(\mathbf{V},\lambda)=
\displaystyle \frac{1}{2} \left| \mathbf{V}-\hat{\mathbf{V}}^{n} \right|^2-\displaystyle\sum_{1\leq i<j\leq N} \lambda_{ij}\, \left( D_{ij}(\mathbf{X}^n) + \Delta t\, \mathbf{G}_{ij}(\mathbf{X}^n)\cdot \mathbf{V}\right).
$$
Notice that the number of Lagrange multipliers corresponds to the number of possible contacts. In particular, if there is no contact between particles $i$ and $j$, then $\Lambda_{ij}=0$ and the Lagrange multiplier is not activated; conversely, if there is a contact between the two spheres, then $\Lambda_{ij}$ may be positive and the corresponding auxiliary field allows the velocity field to satisfy the no-overlapping constraint. The approximate reaction fields $\mathbf{\Lambda}^{n}=(\Lambda_{ij}^{n})$ is the dual component of a solution to the associated saddle-point problem. This problem is solved by an Uzawa algorithm (see, e.g., \cite{CI90}).\\

In the case of ellipsoidal particles, the minimal distance between two ellipses is computed at each time by solving the problem of searching for the proximal points on a each pair of ellipsoidal particles. This problem can be defined in the following way : for all $i<j$, \textnormal{find} $(\x_i,\x_j) \in \R^{2}$ \textnormal{such that}
\begin{align*}
\x_i^n & \in \partial B_i,\\
\x_j^n & \in \partial B_j,\\
\n_i \cdot \n_j & =  - \vert n_j \vert \vert n_j \vert ,\\
\frac{\n_i}{\vert n_i \vert} & =  \frac{\x_i^n \x_j^n}{\vert \x_i^n \x_j^n \vert},
\end{align*}
where $\n_i$ and $\n_j$ are respectively the outward normal to ellipse $B_i$ at point $\x_i^n$ and 
the outward normal to ellipse $B_j$ at point $\x_j^n$. We use a Newton algorithm to numerically solve this problem. The method then consists of projecting the translational but also the angular velocity onto the set of admissible velocities associated to the current configuration.

%The interest in the procedure relies on the possibility to use any suitable solver for the computation of the dynamics. Contacts are handled at a second stage, without any consideration of the proper dynamics. At some point, it allows the use of any solver for the resolution of the dynamics problem: then the so-called predicted velocity field is projected onto the set of admissible velocity fields.

\section{Pair correlations}

\begin{figure}[h]
\includegraphics[width=0.6\textwidth]{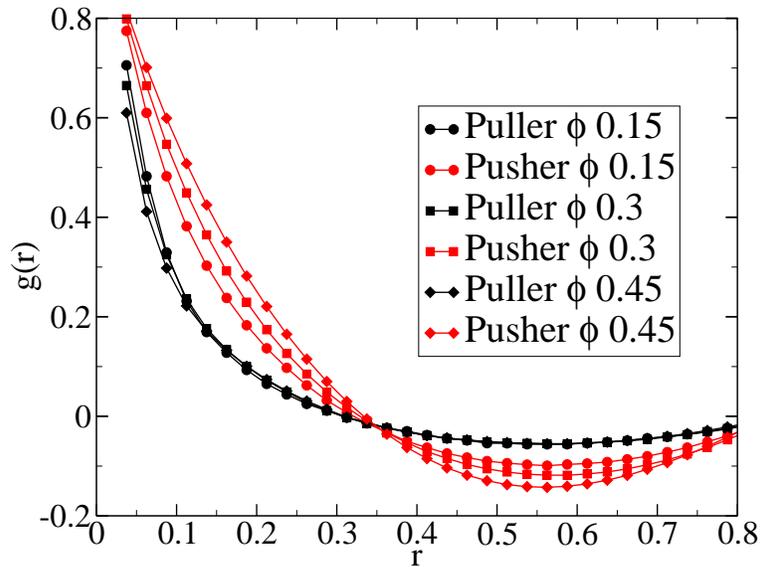}
\caption{Two-point particle correlation computed for active ellipsoidal particles. Pushers are in red and Pullers in black.  }
\label{fig2sup}
\end{figure}
In the main text, we have shown that active particles are able to generate rolls, related to a coherent motion.
In Fig. \ref{fig2sup} we show the pair correlation of the particles, which is the order parameter that measures the level of coherent directional motion of the velocity field \cite{cisneros2011dynamics}.
The field is defined as
\begin{equation}
g(r)=\left \langle \dfrac{{\bf v}({\bf x},t)\cdot{\bf v}({\bf x}+{\bf r},t)} {\vert {\bf v}({\bf x},t)\vert \vert{\bf v}({\bf x}+{\bf r},t)\vert} \right \rangle_{p,t},
\end{equation}
where $\lra{}_{p,t}$ indicates average over time and over all particles.
The pair-correlation points to the local organisation of the system, and indicates that active particles are able to manifest a certain level of coherence up to a certain integral length. The figure explains the qualitative picture discussed in the main text, as Pushers shows a higher degree of organisation up to a larger length, about $0.3L$, whereas Pullers are less correlated and only on a smaller scale, about $0.2L$. Provided hydrodynamic interactions are effective $\phi\gtrsim 0.15$, the effect of the concentration is not strong and clearly negligible for the case of pullers.

\section*{Rheological properties}

\begin{figure}[h]
\includegraphics[width=0.4\textwidth]{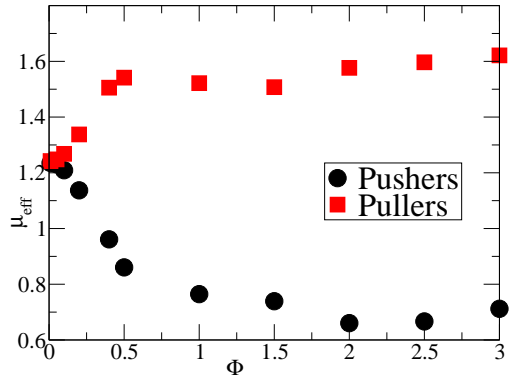}
\caption{Effective viscosity $\mu_{eff}$ with respect to $\Phi$.}
\label{fig4sup}
\end{figure}

The effective viscosity of a suspension of passive spherical particles in a fluid of viscosity $\mu$ depends on its volume fraction $\phi$. In the dilute regime, 
where $\phi \ll 1$ so that particles do not interact, the effective viscosity is well described by Einstein's relation 
%\begin{equation}
%\label{einstein}
$\mu_{\mathrm{eff}} \approx \mu_0 (1 + \alpha \phi),
$
%\end{equation}
where $\alpha$ is known as Einstein's coefficient~\cite{belzon81} and depends on the dimension and on the elongation of entities.  
The linear dependency with respect to volume fraction is due to the fact that, in this regime, the total effect of the particles on the viscosity is equal to the sum of individual contributions. 
Beyond the dilute regime, particles interact and thus simple addition of contributions is no more valid. Polynomial development of the form: $\mu_{\mathrm{eff}} \approx \mu_0 (1 + \alpha \phi + \beta \phi^2 + \dots)$ is needed.
%Work in the semi-dilute regime has focused on finding coefficients for the higher order terms in $\phi$ in a polynomial development of the form: $\mu_{\mathrm{eff}} \approx \mu_0 (1 + \alpha \phi + \beta \phi^2 + \dots)$. 
%In the case of active suspensions, the measurement or computation of the effective viscosity is particularly interesting, since different phenomena can arise depending on the type of motion. 
%Elongated pusher-like swimmers, for instance, tend to decrease the effective viscosity \cite{GLAB11}, while spherical ones have no rheological signature \cite{HABK08,IP07a}.
% On the other hand, experiments in \cite{RJP10} made on the puller-like swimmers \emph{Chlamydomonas} show an increase in the viscosity with respect to the corresponding passive  suspension (measured when the organisms are dead). 
%The different studies have shown that three ingredients are needed in order for self-propelled swimmers to produce a rheological signature: propulsion, elongation and hydrodynamic interactions. Our numerical simulations confirm these results (\emph{faire reference a la figure du texte principal ?}).
%\subsection{Numerical computation of the effective viscosity}
We have computed the effective viscosity in active suspensions of pusher- and puller-like swimmers using 2D simulations in which shear is imposed through non homogenous Dirichlet conditions and periodic boundary conditions are imposed on the left and right boundaries. 
We have considered concentrations up to little less than $45 \%$ volume fraction. 
%Least-squares polynomial regression allows us to fit a polynomial of any degree to numerical data of the form $(\phi^{(1)}, \mu^{(1)})$, $(\phi^{(2)}, \mu^{(2)})$,..., $(\phi^{(N)}, \mu^{(N)})$. Thus we derive the expression of the effective viscosity with respect to the solid fraction $\phi$ as:
%$$
%\mu_{\mathrm{eff.}}\simeq \mu_{0}(1+\alpha \phi+\beta \phi^2+\gamma \phi^3+...).
%$$
%
%Figure \ref{visco:concentration:fit} shows the cubic fit obtained for the effective viscosity with respect to concentration in passive, pusher and puller suspensions.
%%
%\begin{table}[h!]
%\begin{center}
%\begin{tabular}{l|rrr}
%$\eta_{\mathrm{eff.}}$& $\alpha$ & $\beta$ & $\gamma$ \\
%\hline
%{passive} &$+4.3$&$-21.8$ &$+105.6$ \\
%{pushers} &$-5.4$&$+26.2$&$-15.5$ \\
%{pullers} &$+9.8$&$-32.9$&$+170.3$ \\
%\end{tabular}
%\end{center}
%\caption{Coefficients of the mean-least-squares fit describing how the effective viscosity relates to the solid fraction}\label{tab:fit}
%\end{table}
%Table \ref{tab:fit} provides the coefficients of the polynomial of order 3. 
In the main text, we have shown that three ingredients are needed in order for self-propelled swimmers to produce a rheological signature: propulsion, elongation and hydrodynamic interactions.
The explanation has been already given \cite{GLAB11,saintillan2018rheology}, and our numerical simulations naturally retrieve these mechanisms.
%simple explanation can be given for these results.
%  As confirmed by our simulations, in a planar shear flow, 
A single active particle (pusher or puller) rotates with the same angular velocity as a passive particle of same shape, while performing an ellipsoidal trajectory with characteristics depending on its individual velocity and elongation. The contribution to apparent viscosity due to its activity depends on the orientation of the force dipole. 
The contribution of an active particle to apparent viscosity is maximal when the dipole of propulsion forces is against the flow, and it is minimal when this dipole helps the flow. But since the time spent by a particle in each orientation is symmetric with respect to the neutral position (when the dipole is oriented parallel to the walls and apparent viscosity is equal to the passive case), the contribution to viscosity due to the activity vanishes when averaging over one period. Thus we obtain the same effective viscosity as in the passive case. %This is also the case for spherical particles, where the angular velocity is constant. 
In the dilute regime, since effective viscosity is a result of the sum of contributions of each particle, self-propulsion has no impact on the rheology of the suspension. Beyond the dilute regime, hydrodynamic interactions disrupt the rotational behavior of individual swimmers. However, in the absence of a particular alignment of the active particles in the flow, the orientation of the population remains isotropic when averaging in time, and thus the effective viscosity remains the same as in the passive case. In the case of spherical active particles, there are no particular alignments and thus the activity of the particles has no impact on the effective viscosity. This is no longer true when the particles are elongated. 
Indeed, Fig. 3 of the main text shows the probability density function (pdf) of particle orientations obtained for the simulations with spherical and elongated particles, in the passive and in the active case. We see that the symmetry with respect to the neutral orientation is broken in the case of active and elongated particles. Due to interactions with the other particles, the latter tend to align in the flow at an orientation that maximizes their contribution to apparent viscosity in the case of pullers, and minimizes this contribution in the case of pushers. Thus effective viscosity is enhanced in the first case, and diminished in the second case.
%\begin{figure}[h]
%\includegraphics[width=0.6\textwidth]{fig3b.eps}
%\caption{Probability density function of particle orientations. }
%\label{fig3sup}
%\end{figure}
%Coefficient $\alpha$ is particularly meaningful for the moderate concentration regime: when compared to the passive case, the value of $\alpha$ shows that propulsion due to pullers leads to an increase of the effective viscosity. In the case of pushers, propulsion leads to a decrease the effective viscosity. This observation is in good agreement with experiments. When focusing on dense regimes, the first order approximation is not valid anymore and coefficients $\beta$ and $\gamma$ allow us to describe the behavior of the effective viscosity with respect to the solid fraction: in particular, in the case of pushers, the initial decrease of the effective viscosity is counterbalanced by the increase of the solid fraction but the qualitative observation that was formulated in the moderate concentration regime remains valid in the dense regime: the type of motility strongly affects the rheological response of the suspension at the macroscopic level.
We have also investigated how the effective viscosity evolves when the ratio between activity of the micro swimmers and shear rate changes, notably at varying the propulsion activity.
For that purpose we introduce a non-dimensional number that characterizes the shear flow in presence of micro-swimmers :
$
\Phi = \frac{f_p}{\mu S},
$
where $f_p$ is the magnitude of the propulsion force, $\mu$ is the viscosity of the fluid and $S$ the speed of the plates. 
As shown in Fig. \ref{fig4sup},
for low values of the propulsion-shear ratio, the effective viscosity in pusher (resp. puller) suspensions decreases (resp. increases) in a non-linear way.
But above a certain ratio, the effective viscosity stagnates, which is consistent with what found in the literature \cite{GLAB11,lopez2015turning,saintillan2018rheology}.

\end{document}